\documentclass[conference]{IEEEtran}
\usepackage{cite}
\usepackage{amsmath,amssymb,amsfonts}
\usepackage{algorithmic}
\usepackage{graphicx}
\usepackage{subfigure}
\usepackage{textcomp}
\usepackage{xcolor}
\usepackage{bm}
\usepackage{tabu}
\usepackage{booktabs}
\usepackage{diagbox}
\usepackage{bbding}
\usepackage{booktabs}
\usepackage{mathrsfs}
\usepackage{dsfont}
\usepackage{amsfonts}
\usepackage{dsfont}
\usepackage{bm}
\usepackage{textcomp}

\begin{document}

\title{CNN-based Timing Synchronization for OFDM Systems Assisted by Initial Path Acquisition in Frequency Selective Fading Channel}

\author{\IEEEauthorblockN{Chaojin~Qing$^\ast$, Na~Yang$^\ast$, Shuhai~Tang$^\ast$, Chuangui~Rao$^\ast$, Jiafan Wang$^\ast$, and Jinliang Chen$^\ast$}
\IEEEauthorblockA{$^\ast$School of Electrical Engineering and Electronic Information,
Xihua University, Chengdu, 610039, China\\
Email: $^\ast$qingchj@mail.xhu.edu.cn}
}
\maketitle

\begin{abstract}
   Multi-path fading seriously affects the accuracy of timing synchronization (TS) in orthogonal frequency division multiplexing (OFDM) systems. To tackle this issue, we propose a convolutional neural network (CNN)-based TS scheme assisted by initial path acquisition in this paper. Specifically, the classic cross-correlation method is first employed to estimate a coarse timing offset and capture an initial path, which shrinks the TS search region. Then, a one-dimensional (1-D) CNN is developed to optimize the TS of OFDM systems. Due to the narrowed search region of TS, the CNN-based TS effectively locates the accurate TS point and inspires us to construct a lightweight network in terms of computational complexity and online running time. Compared with the compressed sensing-based TS method and extreme learning machine-based TS method, simulation results show that the proposed method can effectively improve the TS performance with the reduced computational complexity and online running time. Besides, the proposed TS method presents robustness against the variant parameters of multi-path fading channels.
\end{abstract}

\section{Introduction}\label{1}
Orthogonal frequency division multiplexing (OFDM) modulation technology has been widely applied in modern wireless communication systems, such as the 5th generation (5G) \cite{b1} and the narrowband-Internet of Things (NB-IoT) \cite{b2}. In OFDM systems, timing synchronization (TS) plays an important role, and its performance seriously affects the subsequent signal processing (e.g., channel estimation \cite{b3}, symbol detection \cite{b4}) and even the performance of the entire communication system \cite{b5}.
However, due to the multi-path interference in wireless communication scenarios, the estimated TS starting point usually appears on the strongest path rather than the first arrival path, which degrades the TS accuracy and thus deteriorates the whole performance of an OFDM system \cite{b6,b7}.

 To enhance the TS performance of the OFDM system against frequency selective fading channels, a series of research achievements have emerged, e.g., \cite{b6,b7,b8,b9,b10,b11,b12,b13,b14,b15,b16}.
The compressed sensing (CS)-based TS is proposed in \cite{b6} to remove multi-path interference by utilizing iterative cancellation to search the first arrival path of the wireless channel.
Although the TS error probability reduces in \cite{b6}, multiple iterations result in high computational complexity and online running time.
In recent years, machine learning (ML) has been introduced into wireless communication technology and achieved good results, such as channel estimation \cite{b8}, channel state information feedback \cite{b9}, and so on.
For TS, extreme learning machine (ELM)-based methods are proposed in \cite{b10,b11,b12,b13} to improve the TS accuracy. In \cite{b10}, the residual TS is investigated by assuming the channel estimation has been achieved.
In \cite{b11}, the ELM network is employed to study the frame synchronization in the scenarios of burst communication systems. To improve the spectral efficiency in the frame synchronization, the ELM-based superposition method is investigated by exploiting inter-frame correlation in \cite{b12}. An ELM-based TS method with label design is proposed in \cite{b13}, which reduces the TS error probability of an OFDM system. Although \cite{b11} and \cite{b12} have improved the frame synchronization performance, the TS of OFDM systems is significantly different from the frame synchronization. Thus, the frame synchronization methods in \cite{b11} and \cite{b12} cannot be directly applied to the TS of OFDM systems.
Despite the ELM-based TS for OFDM systems investigated in \cite{b13}, the challenge of multi-path interference has not been well addressed.
In particular, the ELM network is a feed-forward neural network with a single hidden layer and lacks additional hidden layers to improve its ability of learning features, which limits the accuracy of the ELM-based TS methods \cite{b17}. To improve the accuracy of ELM-based TS methods in OFDM systems, it is of necessity to continuously increase the number of hidden-layer neurons in ELM networks, which raises the computational complexity and brings the risk of overfitting \cite{b17}.

Therefore, to cope with multi-path interference and reduce the computational complexity and online running time, assisted by the initial path acquisition, a convolutional neural network (CNN)-based TS scheme is proposed in this paper.
%
Firstly, the classic cross-correlation method is adopted to estimate the timing offset, and an initial path is captured to assist the following TS.
Based on the captured initial path, a lightweight one-dimensional (1-D) CNN is developed to perform the TS of OFDM systems. On the one hand, the lightweight CNN network replaces the iterative process in [6], reduces the computational complexity and forms a model-driven mode.
On the other hand, the feature extraction ability of CNN surpasses that of the ELM network in \cite{b10,b11,b12,b13} and improves the TS accuracy of OFDM systems against frequency selective fading channels. Simulation results show that, compared with the existing CS-based TS method \cite{b6} and the ELM-based TS method \cite{b13}, the proposed method not only improves the accuracy of TS, but also reduces the computational complexity and online running time.

\textit{Notations:} The notations adopted in this letter are described as follows. $\mathbb{R}^{M\times N}$ and $\mathbb{C}^{M\times N}$ stand for the $M$-by-$N$ dimensional real matrix space and $M$-by-$N$ dimensional complex matrix space, respectively. $[\cdot]^T$, ${\left[  \cdot  \right]^H}$, $\mathrm{E}\{\cdot\}$, and $|\cdot|$ stand for the transpose operation, conjugate transpose operation, expectation operation, and absolute operation, respectively. $[\mathbf{X}]_{m,n}$ denotes the entry $(m,n)$ of matrix $\mathbf{X}$. ${\mathop{\rm Re}\nolimits} \left(  \cdot  \right)$ and ${\mathop{\rm Im}\nolimits} \left(  \cdot  \right)$ represent the operation of taking the real and imaginary parts of a complex value or a complex value matrix, respectively.
\section{System Model}\label{2}
We consider an OFDM system with $N$ subcarriers, which contains additional ${N_g}$ samples for guard interval. The time domain signal is obtained by employing the inverse discrete Fourier transform (IDFT) \cite{b18}, i.e.,
\begin{equation}
\label{eq_1}
\tilde x\left( n \right) = \frac{1}{{\sqrt N }}\sum\limits_{k = 0}^{N-1} {X\left( k \right){e^{j\frac{{2\pi kn}}{N}}}} ,0 \le n \le N - 1,
\end{equation}
where $X\left( k \right)$ stands for the data symbol on the $k$-th subcarrier. Before sending the signal $\tilde x\left( n \right)$, the approach of zero padding (ZP) is utilized to form its protection interval and generate the time domain transmission signal $x\left( n \right), - {N_g} \le n \le N - 1$, with the length of ${N_u}$ (${N_u} = N + {N_g}$), where ${N_u}$ represents the length of the observation window.
The impulse response of multi-path fading channels is given by\cite{b19}
\begin{equation}
\label{eq_2}
\ h\left( n \right) = \sum\limits_{l = 0}^{L - 1} {{h_l}\delta \left( {n - {\tau _l}} \right)},
\end{equation}
where $L$ is the number of distinguishable paths, ${h_l}$ and ${\tau _l}$ denote the complex gain and path delay of the $l$-th path, respectively \cite{b11}.
After the transmitted signal $x\left( n \right)$ transmits over multi-path channels, the received time domain signal $y\left( n \right)$, can be expressed as \cite{b20}
\begin{equation}
\label{eq_3}
y\left( n \right) = \sum\limits_{l = 0}^{L - 1} {{h_l}x\left( {n - \tau  - {\tau _l}} \right) + w\left( n \right)},
\end{equation}
where $\tau $ represents the timing offset to be estimated, and $w\left( n \right)$ is the complex additive white Gaussian noise with zero mean and variance ${\sigma ^2}$.
\section{Timing Synchronization Scheme}
\label{Section_III}
\begin{figure*}[t]
  \centering
    \vspace{-3mm}
  \includegraphics[width=1.6\columnwidth]{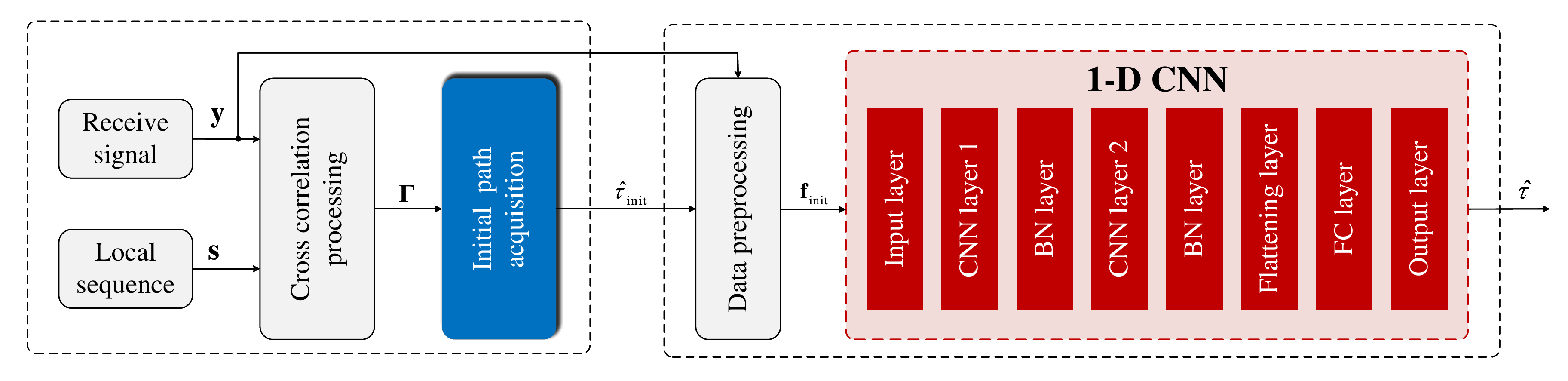}\\
  \caption{Flow chart of the proposed TS method.}
  \label{fig1}
\end{figure*}

To alleviate the multi-path interference, a CNN-based TS scheme for OFDM systems assisted by the initial path acquisition is proposed in this paper, whose flow chart is shown in Fig. \ref{fig1}.
We first present the initial path acquisition in \textit{Section \ref{subsection31}}. Then, in \textit{Section \ref{subsection_3_2}}, the CNN-based TS scheme is elaborated. Finally, the computational complexity and online running time are discussed in \textit{Section \ref{4_AC}}.


\subsection{Initial Path Acquisition}\label{subsection31}

In this paper, the classical cross-correlation method \cite{b21} is used for capturing an initial path. According to (\ref{eq_3}), we observe $M$ received samples, where $M = N + {N_g} + {N_d}$ with $N_d$ being the length of transmitted data. By denoting the $M$ received samples as ${\bf{y}} \in \mathbb{C}{^{M \times 1}}$, i.e., ${\bf{y}} = {\left[ {y\left( 0 \right),y\left( 1 \right), \cdots ,y\left( {M - 1} \right)} \right]^T}$,
the timing metric, denoted by $\bf{\Gamma }$, can be expressed as:
\begin{equation}
\label{eq_5}
{\bf{\Gamma  = }}{{\bf{S}}^H}{\bf{y}},
\end{equation}
where ${\bf{\Gamma }} \in \mathbb{C} {^{{N_{{\rm{lag}}}} \times 1}}$ with ${N_{{\rm{lag}}}}$ ($N_{{\rm{lag}}} \le M - N$) being the length of the timing metric vector.
${\bf{S}} \in \mathbb{C} {^{M \times {N_{{\rm{lag}}}}}}$ is a cyclic shift matrix formed according to the cyclic shift of the training sequence \cite{b11}, i.e.,
\begin{equation}
\label{eq_6}
{\bf{S}} = \left[ {\begin{array}{*{20}{c}}{{s_0}}&0& \cdots \\\vdots &{{s_0}}& \ddots \\{{s_{N - 1}}}& \vdots & \ddots \\0&{{s_{N - 1}}}& \ddots \\ \vdots &0& \ddots \\ {}& \vdots & \ddots \end{array}} \right],
\end{equation}
where the first column of ${\bf{S}}$ is formed by the training sequence ${\bf{s}} = {\left[ {{s_0},{s_1}, \cdots ,{s_{N - 1}}} \right]^T}$. Then, the initial auxiliary point of TS, denoted by ${\widehat \tau _{{\rm{init}}}}$, is given by
\begin{equation}
\label{eq_7}
{\widehat \tau _{{\rm{init}}}} = \mathop {\arg \max }\limits_{0 \le d \le {N_{{\rm{lag}}}} - 1} \left\{ {\left| {\Gamma \left( 0 \right)} \right|, \cdots ,\left| {\Gamma \left( {{N_{{\rm{lag}}}} - 1} \right)} \right|} \right\},
\end{equation}
where $\Gamma \left( d \right)$, $0 \le d \le {N_{{\rm{lag}}}} - 1$, is the $d$-th element of ${\bf{\Gamma }}$.

According to \cite{b6}, with the achieved ${\widehat \tau _{{\rm{init}}}}$, the TS point relative to the first arrival path is restricted to the range $\left[ {\max \left( {{{\widehat \tau }_{{\rm{init}}}} - {N_g},0} \right),{{\widehat \tau }_{{\rm{init}}}}} \right]$. Then, the search range of the TS point is narrowed, which can simplify the followed 1-D CNN and thus reduces its computational complexity and online running time.



\subsection{CNN-based TS}\label{subsection_3_2}
In the cross-correlation based TS \cite{b21}, the strongest path rather than the first arrival path is usually searched as the TS point, which is vulnerable to the multi-path interference.
To tackle this issue, the initial path acquisition is employed to narrow the search range of TS for the following 1-D CNN. Then, we develop a 1-D CNN to simplify the iterative interference cancellation in \cite{b6}, or the ELM networks in \cite{b10,b11,b12,b13}. To this end, the multi-path interference is alleviated and thus paves the way for TS to search the first arrival path.


\subsubsection{Network Architecture}\label{subsubsection321NS}
 To reduce the computational complexity and online running time, we first employ 1-D CNN to construct the TS network. As summarized in \textit{TABLE \ref{table_I}}, this 1-D CNN is optimized by specially designing its filter size (or convolution kernel size) and filter number.
The main idea is that the TS accuracy is closely related to timing metrics \cite{b13}, and our main considerations are as follows.

%
\begin{itemize}
\item In the first convolution layer, a convolution kernel with a length of $2N$ forms a relatively large receptive field (relative to the training sequence with the length of $N$). Through the first convolution layer, the complete features of the timing metrics formed by the training sequence are captured from the input signal.
\item The second convolution layer uses a relatively small convolution kernel size to capture significant features about the channel (e.g., power delay profile (PDP)) from the complete features, whose mechanism originates from the threshold-based detection method for the first arrival path \cite{b6,b22}. The difference is that 1-D CNN is employed to cancel the multi-path interference rather than the mode of iterative interference cancellation in \cite{b6}.
\end{itemize}

\begin{table}
\renewcommand{\arraystretch}{1.25}
\caption{Network Architecture}
\label{table_I}
\centering
\scriptsize
\setlength{\tabcolsep}{0.5mm}{
\begin{tabu}{@{}c|c|c|c|c@{}}
\tabucline[1pt]{-}
    Layer Name       & Size of output  &  Size of filter &Number of filter & Activation\\ \tabucline[1pt]{-}
    Input layer & $2\left( {{N_u} - 1} \right) \times 1\times 1$ & - & - & - \\ \hline
    CNN Layer 1 & $\left( {2{N_g} - 1} \right) \times 4\times 1$ & $2N\times 1$ & 4 & - \\ \hline
    BN Layer & $\left( {2{N_g} - 1} \right) \times 4\times 1$ & - & - & ReLU \\ \hline
    CNN Layer 2 & ${N_g} \times 4\times 1$ & ${N_g}\times 1$ & 4 & - \\ \hline
    BN Layer & ${N_g} \times 4\times 1$ & - & - & ReLU \\ \hline
    Flattening & $4{N_g} \times 1$ & - & - & - \\ \hline
    FC Layer & ${N_g}\times 1$ & - & - & Sigmoid \\ \hline
    Output Layer & ${N_g}\times 1$ & - & - & Softmax \\
    \tabucline[1pt]{-}
\end{tabu}}
  \vspace{-3mm}
\end{table}

\subsubsection{Offline Training}\label{subsubsection322OT}
We first describe the collection of training data set $\left\{ {{{\bf{f}}_{{\rm{init}},i}}}, {{{\bf{t}}_i}} \right\}_{i = 1}^{{N_t}} $ with ${N_t}$ samples, where ${{\bf{f}}_{{\rm{init}},i}} \in \mathbb{R} {^{2{N_u} \times 1}}$ and ${{\bf{t}}_i} \in \mathbb{R} {^{{N_g} \times 1}}$ are the training data and learning label of the $i$-th sample, respectively. According to (\ref{eq_1})--(\ref{eq_3}), the received signal ${{\bf{y}}_i}$ is obtained.
Then, the initial auxiliary point of TS ${\widehat \tau _{{\rm{init,}}i}}$ is captured according to (\ref{eq_5})--(\ref{eq_7}).
According to ${{\bf{y}}_i}$ and ${\widehat \tau _{{\rm{init}},i}}$, the observation signal ${{\bf{r}}_i} \in \mathbb{C} {^{({N_u}-1) \times 1}}$ is given by
\begin{equation}
\label{eq_9}
{{\bf{r}}_i} = \left\{ {\begin{array}{*{20}{l}}
{{{\left[ {{y_{i,{{\widehat \tau }_{{\rm{init}},i}} - {N_g} + 1}}, \cdots ,{y_{i,{{\widehat \tau }_{{\rm{init}},i}} + N - 1}}} \right]}^T}}&{,{\widehat \tau _{{\rm{init}},i}} \in {\Omega _1}}\\
{{{\left[ {{y_{i,0}},{y_{i,1}}, \cdots ,{y_{i,N + {N_g} - 1 - 1}}} \right]}^T}}&{,{\widehat \tau _{{\rm{init}},i}} \in {\Omega _2}}
\end{array}} \right.,
\end{equation}
where ${\Omega _1}$ and ${\Omega _{{\rm{2 }}}}$ represent the intervals of ${\widehat \tau _{{\rm{init}},i}} \ge {N_g} - 1$ and ${\widehat \tau _{{\rm{init}},i}}{\rm{ < }}{N_g} - 1$, respectively. Usually, the real-valued training data is required by a neural network \cite{b23}. Thus, as shown in Fig. \ref{fig1}, the complex-valued ${{\bf{r}}_i}$ is transformed into the real-valued training data ${{\bf{f}}_{{\rm{init}},i}}$ according to
\begin{equation}
\label{eq_10}
{{\bf{f}}_{{\rm{init}},i}} = {\left[ {{\mathop{\rm Re}\nolimits} \left( {{r_{i,0}}} \right),{\mathop{\rm Im}\nolimits} \left( {{r_{i,0}}} \right), \cdots ,{\mathop{\rm Re}\nolimits} \left( {{r_{i,{N_u} - 2}}} \right),{\mathop{\rm Im}\nolimits} \left( {{r_{i,{N_u} - 2}}} \right)} \right]^T}.
\end{equation}
where $r_{i,j}$ with $0 \leq j \leq {N_u} - 2$ is the $j$-th entry of ${{\bf{r}}_i}$.

%
%
%
According to the initial auxiliary point ${\widehat \tau _{{\rm{init}},i}}$ and the timing offset ${\tau _i}$, by using one hot coding, the learning tag ${{\bf{t}}_i} \in \mathbb{R} {^{{N_g} \times 1}}$ is constructed as \cite{b24}
\begin{equation}
\label{eq_11}
{{\bf{t}}_i} = {\left[ {\underbrace {0, \cdots ,0}_{{\tau _{r,i}}},1,\underbrace {0, \cdots ,0}_{{N_g} - {\tau _{r,i}} - 1}} \right]^T},
\end{equation}
where ${\tau _{r,i}}$ ($0 \le {\tau _{r,i}} \le {N_g} - 1$) represents the timing offset of the first arrival path of the $i$-th training sample. Based on the initial auxiliary point of TS, i.e., ${\widehat \tau _{{\rm{init}},i}}$, we have
\begin{equation}
\label{eq_12}
{\tau _{r,i}} = {\widehat \tau _{{\rm{init}},i}} - {\tau _i}.
\end{equation}
With the training data set $\left\{ {{{\bf{f}}_{{\rm{init}},i}}}, {{{\bf{t}}_i}} \right\}_{i = 1}^{{N_t}} $, the 1-D CNN is trained to optimize its network parameters \cite{b25}.

\subsubsection{Online Deployment}\label{Online_Deployment}
First, the initial auxiliary point of TS, i.e., ${\widehat \tau _{{\rm{init}}}}$, is obtained according to (\ref{eq_1})--(\ref{eq_7}). With the assistance of ${\widehat \tau _{{\rm{init}}}}$, ${{\bf{f}}_{{\rm{init}}}}$ is gained by using (\ref{eq_9}) and (\ref{eq_10}). Then, the network output of 1-D CNN, denoted as $ {\bf{O}} \in \mathbb{R}^{^{{N_g} \times 1}} $, is obtained according to the network input ${{\bf{f}}_{{\rm{init}}}}$. By denoting $ {\bf{O}}$ as ${\bf{O}} = {\left[ {{O_0},{O_1}, \cdots ,{O_{{N_g} - 1}}} \right]^T}$, the timing offset relative to ${\widehat \tau _{{\rm{init}}}}$, denoted as ${\widehat \tau _r}$, is given by
\begin{equation}
\label{eq_18}
{\widehat \tau _r} = \mathop {\arg \max }\limits_{0 \le j \le {N_g} - 1} \left\{ {{O_j}} \right\}.
\end{equation}
According to ${\widehat \tau _{{\rm{init}}}}$ and ${\widehat \tau_r}$, the estimated timing offset ${\widehat \tau}$ of the OFDM system is
\begin{equation}
\label{eq_19}
\widehat \tau  = {\widehat \tau _{{\rm{init}}}} - {\widehat \tau _r}.
\end{equation}

\subsection{Computational Complexity and Online Running Time}\label{4_AC}


\begin{table*}
  \vspace{-3mm}
\renewcommand{\arraystretch}{1.25}
\caption{Computational Complexity and Online Running Time among Different TS Methods}
\label{table_II}
\centering
\scriptsize
\setlength{\tabcolsep}{0.8mm}{
\begin{tabu}{@{}c|c|c|c@{}}
\tabucline[1pt]{-}
 Method      &Computational Complexity (Expression)  &Computational Complexity (Example) &Online Running Time (sec)\\
 \tabucline[1pt]{-}
    CS\cite{b6}     & ${6{N_{{\rm{lag}}}}N + \sum\limits_{l = 1}^6 {2lM + 2{l^2}M + {l^3}} }  $             &  $203193$  & $39.5$            \\ \hline
    ELM\cite{b13}   & ${N_{{\rm{lag}}}}N + 20N_{{\rm{lag}}}^2$              &  $541440$ & $16.28$        \\ \hline
    {Proposed}      &         ${N_{{\rm{lag}}}}N + \frac{1}{4}\left( {\sum\limits_{l = 2}^{{\lambda _c}} {{K_l}{N_l}{C_{l - 1}}{C_l}} } \right.\left. { + \sum\limits_{l = 2}^{{\lambda _d}} {{N_{l - 1}}{N_l}} } \right)$&  $44544$   & $6.89$           \\
    \tabucline[1pt]{-}
\end{tabu}}
  \vspace{-3mm}
\end{table*}


The computational complexity (the complex multiplication is applied as the metrics of computational complexity) and online running time among different methods for TS are listed in \textit{TABLE \ref{table_II}}, where ${K_l}$ represents the size of convolution kernel; ${N_l}$ and ${N_{l - 1}}$ stand for the network outputs of the $l$-th layer and the ($l - 1$)-th layer, respectively; ${C_l}$ and ${C_{l - 1}}$ are the number of convolution channels of the $l$-th and ($l-1$)-th convolution layers, respectively; ${\lambda _c}$ and ${\lambda _d}$ delegate the total number of layers of convolution layer and full connection layer, respectively. An example is also given in \textit{TABLE \ref{table_II}}, where $N = 128$, ${N_g} = 32$, $L = 20$, $M = 288$, and $\eta  = 0.2$ (the channel exponentially decayed factor) are considered. Then, ${N_{{\rm{lag}}}} = M - N + L = 180$. In addition, $10^4$ simulations are employed to evaluate the online running time, and the iterations in \cite{b6} are set as 6.



It can be seen from the example that, compared with the CS-based TS method \cite{b6} and the ELM-based TS method \cite{b13}, the proposed TS method of this paper reduces the computational complexity and online running time. This mainly benefits from the assistance of initial path acquisition and the construction of lightweight network architecture. Furthermore, relative to the methods in \cite{b6} and \cite{b13}, the proposed method improves the TS accuracy with reduced computational complexity and online running time, which will be elaborated in \textit{Section IV}.
\section{Numerical Simulation}
\label{section_IV}

%

In this section, we evaluate the TS performance according to the TS error probability. The parameters involved in the simulations are given as follows. $N = 128$, ${N_g} = 32$, and $M = 288$ are considered. ${N_u} = N + {N_g} = 160$. Zadoff-Chu sequence \cite{b21} is employed as the training sequence.
 %
%
%
 For the convenience of expression, we employ ``Prop'', ``Ref\_[6]'', and ``Ref\_[13]'' to denote the proposed TS method, the CS-based TS method in \cite{b6}, and the ELM-based TS method in \cite{b13}, respectively.

\begin{figure}[t]
  \centering
  \vspace{-3mm}
  \includegraphics[width=0.4\textwidth]{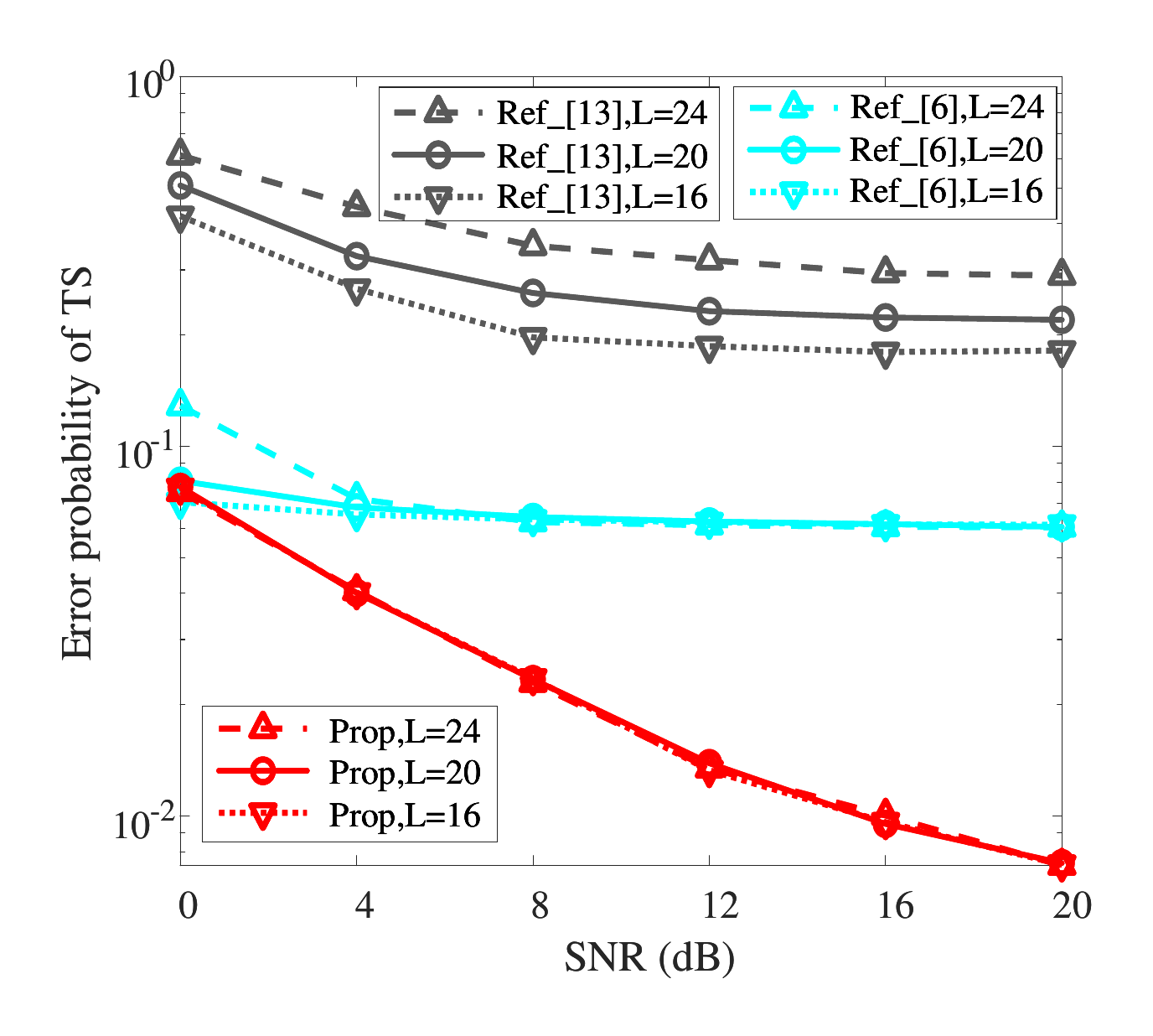}\\
  \caption{The robustness analysis of $L$.}
  \label{fig2}
\end{figure}

\begin{figure}[t]
  \centering
  \vspace{-3mm}
  \includegraphics[width=0.4\textwidth]{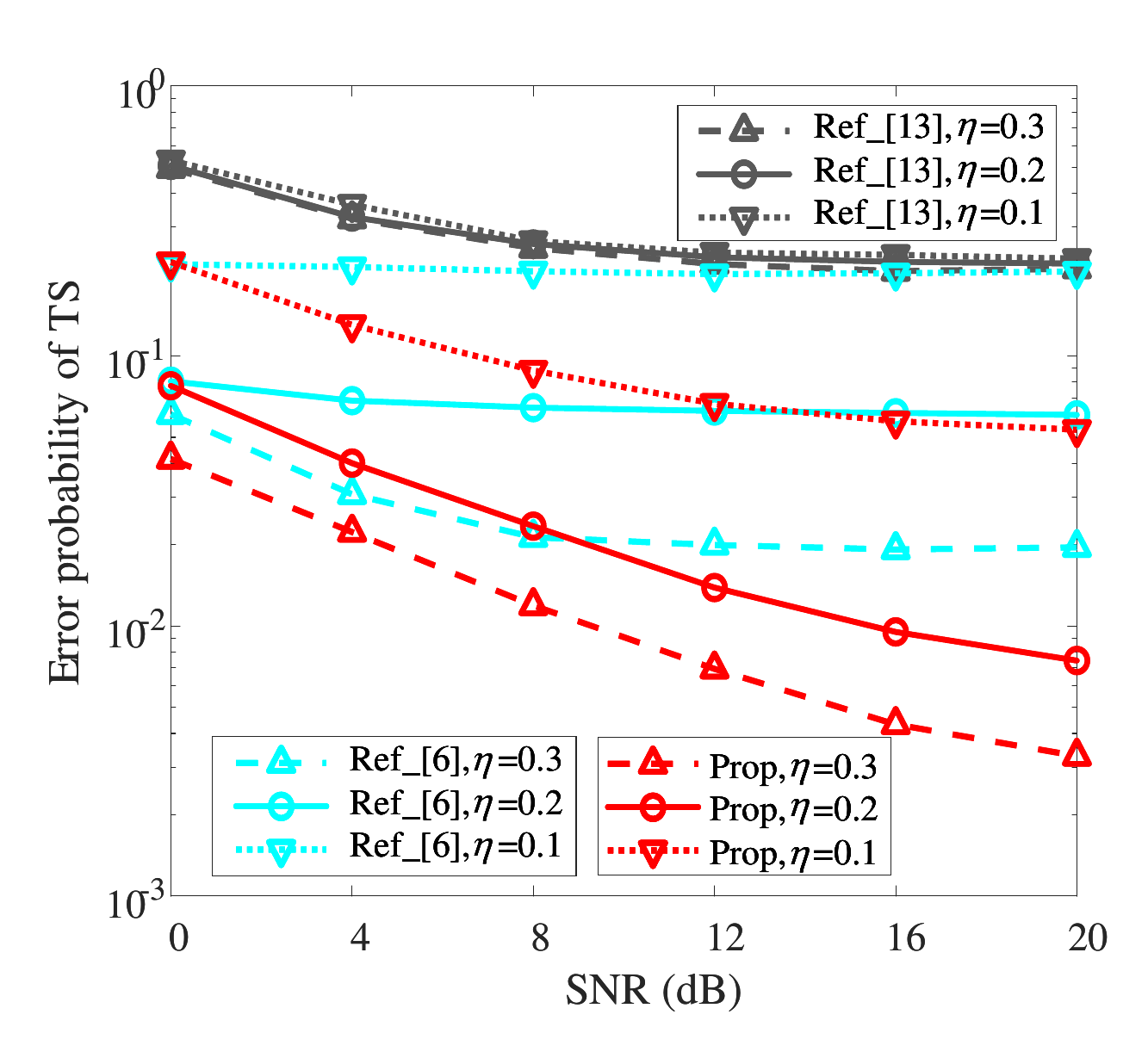}\\
  \caption{The robustness analysis of $\eta$.}
  \label{fig3}
\end{figure}

We use the case where $L = 20$ and $\eta  = 0.2$ \cite{b11} to evaluate the effectiveness of the proposed TS method, which is given in Fig. \ref{fig2}.
From Fig. \ref{fig2}, the ``Prop'' achieves the minimum TS error probability. Especially, for relatively high SNR regions (e.g., SNR $ > $ 8dB), the error probability of TS is distinctly smaller than those of ``Ref\_[6]'' and ``Ref\_[13]''. For example, when SNR = 12dB, the TS error probabilities of ``Ref\_[13]'' and ``Ref\_[6]'' are about 0.23 and 0.06, respectively, while the TS error probability of ``Prop'' is only 0.01. Therefore, compared with ``Ref\_[13]'' and ``Ref\_[6]'', the case where $L = 20$ and $\eta  = 0.2$ \cite{b11} reflects that the ``Prop'' effectively improves the TS accuracy of OFDM systems.

To demonstrate the robustness of the proposed TS scheme against the impact of $L$, the error probabilities of TS for the case where $L=16$, $L=20$, and $L=24$ are also plotted in Fig. \ref{fig2}.
Form Fig. \ref{fig2}, relative to the ``Ref\_[13]'' and ``Ref\_[6]'', the ``Prop'' obtains the minimum error probability of TS when SNR $ > $ 4dB, which demonstrates that ``Prop'' improves the TS's accuracy of the existing methods (e.g., ``Ref\_[6]'', and ``Ref\_[13]'') with the variations of $L$.
With the increase of $L$, the TS error probabilities of ``Ref\_[6]'', and ``Ref\_[13]'' increase for each given SNR due to the increased multi-path interference. For the ``Prop'', however, only a slight fluctuation of TS error probability is observed for a given SNR. This reflects the ``Prop'' has its robustness against the impact of multi-path interference. Especially, for each given value of $L$,
the ``Prop'' obtains the smallest error probability of TS as SNR $\geq $ 4dB. In relatively high SNR region, e.g., SNR $ > $ 12dB, the TS error probability of ``Prop'' is significantly smaller than those of ``Ref\_[6]'', and ``Ref\_[13]''. Therefore, against the impact of $L$, the ``Prop'' reduces the TS error probabilities of ``Ref\_[6]'', and ``Ref\_[13]'', especially in the relatively high SNR region.

Besides, to verify the robustness of the proposed TS scheme against the impact of $\eta$, the error probabilities of TS are illustrated in Fig. \ref{fig3} with different values of $\eta$ (i.e., $\eta = 0.1$, $\eta = 0.2$, and $\eta = 0.3$ \cite{b26} are considered).
For the given values of SNR (from 0dB to 20dB), no matter what the value of $\eta$, the ``Prop'' obtains almost the lowest error probability among the given TS methods. For the case where SNR = 16dB and $\eta = 0.3$, the TS error probabilities of the ``Prop'', ``Ref\_[6]'', and ``Ref\_[13]'' are about 0.004, 0.02, and 0.21, respectively. This illuminates that the ``Prop'' is robust against the impact of $\eta$. For a given SNR, with the increase of $\eta$, the error probabilities of all TS methods decrease due to the more significant difference in PDP. Even so, compared with ``Ref\_[6]'' and ``Ref\_[13]'', the ``Prop'' achieves smaller error probability of TS (especially in relatively high SNR regions, e.g., SNR $\geq4$dB), and thus can effectively reduce the error probability of TS.


On the whole, with the reduced computational complexity and online running time, the ``Prop'' reduces the TS error probability compared with ``Ref\_[6]'' and ``Ref\_[13]'', which reflects its  effectiveness. Against the impacts of $L$ and $\eta$, the simulation results demonstrate that the ``Prop'' has its robustness.

  \vspace{-2mm}
\section{Conclusion}
In this paper, a CNN-based TS scheme is proposed to optimize the OFDM system affected by frequency selective fading channels. In the proposed scheme, the classic cross-correlation is applied to capture the initial path, which is vital to design a specific 1-D CNN network. With the assistance of initial path acquisition, the search region of TS shrinks, and the following 1-D CNN architecture is lightweight. Compared with the CS-based TS method and the ELM-based TS method, the proposed method can effectively improve the TS accuracy with reduced computational complexity and online running time. And its robustness is validated by the stable performance against parameter variations.

  
\end{document}